\documentstyle[NATO]{crckapb}

\newcommand{\bi}{\bibitem}
\newcommand{\PRB}{Phys. Rev. B }
\newcommand{\be}{\begin{equation}}
\newcommand{\ee}{\end{equation}}
\newcommand{\nn}{\nonumber}
\newcommand{\bea}{\begin{eqnarray}}
\newcommand{\eea}{\end{eqnarray}}
\input{epsf}
\begin{opening}
\title{Sub-wetting layer continuum states
  in quantum dot samples}
\author{K. Kr\' al}
\author{P. Zden\v ek}
\institute{Institute of Physics, Academy of Sciences of Czech
Republic\\Na Slovance 2, 18221 Prague 8, Czech Republic}
\end{opening}
\runningtitle{Sub-WL continuum}
\begin{document}

\begin{abstract}
In the polar semiconductor samples of the self-organized quantum dots,
grown by the Stranski-Krastanow growth method, the lowest energy
extended states of the electronic excitations are assumed to be the
wetting-layer states. The coupling between these extended
states and the electronic states localized in the individual quantum
dots, may influence the optical spectra of such samples in the
sub-wetting layer region of energy. This effect is studied
assuming  the Fr\"ohlich's coupling between the electrons
and polar optical phonons. The contribution of this
 interaction to the appearence of the sub-wetting layer
continuum in the optical spectra
and to the level broadening of the localized states, pointed
out in some experiments, is estimated.
\end{abstract}
\section{Introduction}
  The semiconductor nanoparticles are promissing from the point of view
of both basic science and applications \cite{Yoffe}.
One of their significant features is
 that the charge carriers can be localized within these
nanoparticles. The quantum dots
are thus  regarded as a  realization of artificial atoms with the
nearly delta function-like spectral density of the electronic states.
The quantum dot aggregates are prepared very often by the
Stranski-Krastanow growth technique \cite {Yoffe}. In the
self-assembled quantum dot (SAQD) samples, prepared by this method,
the quantum dots are grown on the top of the wetting layer (WL), with
which the quantum dots  interact. The optical
absorption spectra of nanoparticles dispersed in a polymer film,
which also seem to display  an absorption continuum background
increasing with  the energy
of the absorbed photon   \cite{Mittleman,Banin}, are not discussed here.

\begin{figure}
\epsfysize=7cm
\begin{center}
\epsffile{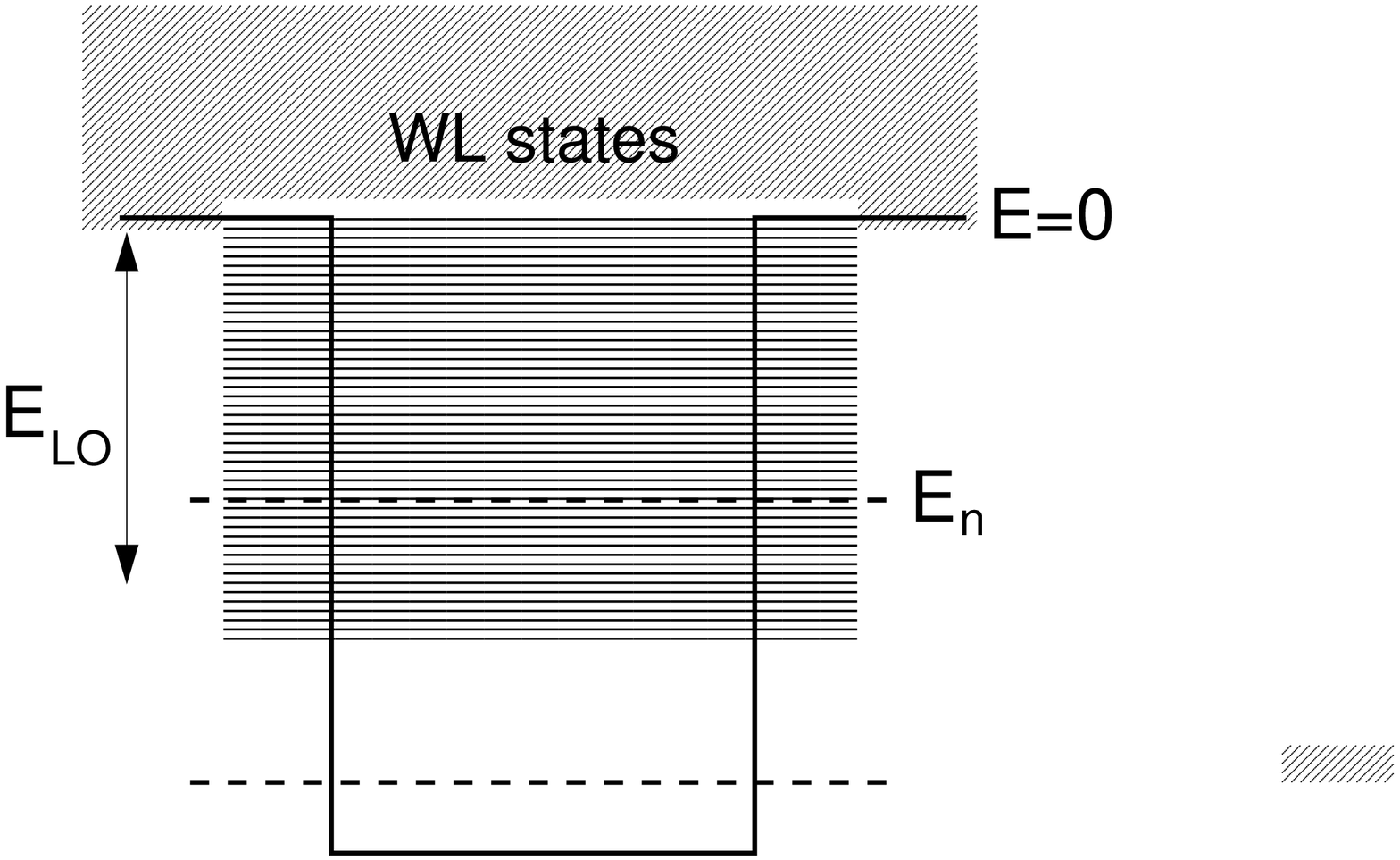}
\end{center}
\caption{Schematic picture of the energetic structure of the single-electron states
in the self-assembled quantum dot (SAQD) sample. The full line is the
potential profile of the electrons in the quantum dot.
The dashed horizontal lines denote
the states bound in the quantum dots. The oblique hatched area denotes the
extended electronic states in the wetting-layer. The horizontal hatching marks the
sub-wetting layer area of electronic energies. $E_n$ is the energy of a selected
bound state in the sub-wetting layer energy region, $E_{LO}$ is optical phonon energy.}
\label{schema}
\end{figure}

 In the type {\it I} quantum dots \cite{Yoffe} in the undoped SAQD
samples, the lowest energy  excitations are the states with the
electrons excited over the semiconductor band gap to the conduction
band electronic quantum dot localized states, leaving the holes in the
valence-band states of the quantum dot. In the SAQD
samples it is expected that the lowest energy electronic excitations,
which are not localized in the quantum dots, are the excitations to
the wetting layer states \cite{Toda,Hinooda}.  The scheme of
the electronic states available to the electronic excitations in the SAQD samples is
drawn in the Fig.\ \ref{schema}.

 The electronic system
can also be excited to the extended states belonging to the substrate
and to the electronic states in the cladding layers
\cite{Schmidt}.
With a certain simplification, the electronic states localized in the quantum
dots can be expected to have a delta function like density of states.
This may be seen  in contrast with
   experiments
\cite{Toda,Schmidt,Steer,Hessman,Heitz,Ledentsovftp,Sauvage,Lemaitre,Finley,Nakaema}
showing that
the spectral densities may be different from delta functions
and that the width of them may depend on the intensity of laser light excitating
the SAQD sample. Besides this observation,  the experiments on
luminescence
and optical absorption  suggest that in the sub-wetting layer energy
region, where we expect only bound electronic
 states, a continuum background is observed.

   The continuous background, which seems  characteristic of the SAQD
samples, occurs often in the  measurements on samples with a rather
high density of the quantum dots and  a large density of electrons
excited to the WL states \cite{Nakaema}. The presence of high densities
of charge carriers may lead us to consider the electrostatic coupling
among the charge carriers \cite{Gelmont,Kempa,Morris}. We shall not consider
this mechanism in the present work. Also, we shall neglect any direct
coupling between quantum dots, like the electron tunneling
\cite{double}, electrostatic coupling between the dots, or any other.
Recently, the Fr\"ohlich's coupling \cite{Callaway}
between the electrons and dispersionless
polar optical phonons has been applied to the
interpretation of the energy relaxation of electrons in  quantum dots
\cite{jarelax,tsuchiya} and to other properties
\cite{jarelax,jaspectrdens,mnichov}. The theory of the energy
relaxation in the two level system with the dispersionless phonons has
recently been supported by an exact calculation  \cite{Mensik}.
The purpose of this work is to show the possible contribution of the
Fr\"ohlich's coupling to the origin of the sub-WL continuous background
spectral density and to the  electronic level broadening in
the SAQD samples.

\section{The theoretical model of the SAQD system}
Trying to study  the optical spectra in the rather complicated samples, as
the SAQD are, we are led to make a number of simplifications, which
make the problem tractable, but which  hopefully do not
 influence principally  the main conclusions of the work.
Dealing with the electron-phonon Fr\"ohlich's interaction
we shall neglect the influence of the interfaces in the SAQD heterostructure
 on the structure
of the optical phonons.  From the reasons given e. g. in the
ref. \cite{Rucker}
we shall simply assume that the charge carriers interact with the dispersionless
longitudinal phonons of the bulk crystal of GaAs.
Neglecting the effect
of the electrostatic interaction among electrons and holes,
we neglect also the exciton effect.
 We shall assume the limit of very heavy
holes, in which their influence is reduced  only to  the static
electrostatic field contributing to the effective potential in which
the electrons move in the quantum dot potential in the conduction band.

In order that the details of the present model Hamiltonian are better understood,
let us remind the
principal features of the structure of the SAQD sample.
With a certain simplification, in the Stranski-Krastanow method the surface of a substrate
is first covered by a thin layer of the material, called wetting-layer,
from which the quantum dots are
to be grown. After depositing this layer the sample is  left to obey a spontaneous
process. In this process the wetting layer changes its structure, namely, due to the
difference in the lattice constant between the substrate material and that
of the  wetting layer,
the  wetting layer starts to become thinner and the material of the wetting layer forms small
nanocrystals on the top of the wetting-layer.
These small nanostructures may have sometimes a shape of a pyramid.
Because of the shape and the size these small structures are called quantum dots.
The process stops at some size of the nanocrystals and at some remaining
thickness of the wetting-layer. The final state is a result of an enegretic equilibrium
in the whole system of the interatomic forces of the system. In order to prevent the mutual
transfer of the matterial between the wetting layer and the quantum dots after
the growth is ended,
all the structure is covered from the top by a capping layer of
a material, which can be similar to the substrate material.
In this structure the electrons appear to be localized in the quantum dots. In order
to overcome this localization they either have to gain an energy to transfer to the capping
layer, or they need to get an energy sufficient for to transfer to the lowest
energy states in the very thin weting layer. Because of the small thickness of
the wetting layer, the lowest energies in this structure are rather large. Usually however,
the lowest energetic barrier for the electron to leave the quantum dot is that
to the wetting layer.
The density of the    quantum dots, with which they are distributed on the surface
of  WL, depends on the technological conditions of the Stranski-Krastanow process.

The wetting-layer is a quasi-two dimensional quantum well nanostructure.
In the SAQD
sample the electrons of the quantum dots are expected to be
 coupled to electronic states
of the wetting layer. The complete orthonormal set of the electronic
states in the quantum dots and in the wetting layer depend on the
details of the geometry of the SAQD sample.
Realizing that besides the WL extended states there
are also other extended states in the real sample,  namely those
associated with the substrate and capping layers \cite{Toda},
which may be energetically not very far from   the wetting-layer states,
and
in order to simplify the theory, we shall assume that
the wetting layer is a three-dimensional subsystem
with the electronic states extended throughout the whole SAQD sample.
The impact of
this dimensionality assumption can be verified by recalculating the
present results
with  a two-dimensional wetting layer assumption. This will be postponed
to a further work. Our model, using the three-dimensional wetting-layer
continuum, may therefore correspond better to the vertically coupled
three-dimensional stacks of the quantum dots \cite{Ledentsov3dstacks},
or, to the Volmer-Weber island growth technology without the wetting layer
 (see e. g.
\cite{Matsumura}).

We shall approximate the electronic bound states in the individual dots by the
electronic eigenstates
in the three dimensional infinitely deep quantum wells.
 Although we shall assume that the density
of the quantum dots in the sample is rather high,
namely 8$\times 10^{21}$\,m$^{-3}$, which by the mean inter-dot
distance corresponds to the usual two-dimensional density
of quantum dots in the measured samples \cite{Finley},
we shall assume that the single electron states in the individual
quantum dots are mutually orthogonal. In this work,
confining ourselves to the basic estimates,
 we shall need to consider only
one bound state per a single dot.
In real samples, the individual quantum dots have
only a finite potential depth.   In our model we shall therefore
 assume that the single-electron spectrum in a quantum dot
is shifted by a certain energy, so that the energy of the
bound state under consideration is positioned in the sub-wetting layer
region of electronic energy and we shall neglect the other bound states in the dots.

The electronic motion in the WL states can be approximated by plane waves,
with the electronic energy $E_{\bf k}=\hbar^2 \mid {\bf k} \mid^2 /(2m )$,
$m$ being the
electron conduction band energy in GaAs.
The plane waves together with the bound states would not make an
orthogonal set of states. The problem of
nonorthogonality of the basis can be approximately
avoided upon restricting the magnitude $k$ of
the electron wave vectors ${\bf k}$
in the WL states to the values $k \ll k_m$. In this work we shall assume that
the electronic wavelength  corresponding to $k_m$ is $1.5*d$,
$d$ being the lateral size of the cubic quantum dot. We assume $d=20$\,nm.
In GaAs this assumption limits the WL electronic energy from above by about
24\,meV.

The restriction put on the extent of the WL space of states will be utilized
 also in another context. We know little about the state of
electronic statistical distribution in the course of the experiment, so that we
use a simple assumption of the population of the WL states.
  Namely, we shall assume that the
quantum mechanical electronic WL states are
 populated by electrons with a homogeneous density $\overline{N_{\bf k}}$, independent of
 the electron wave vector.
 This homogeneity assumption simplifies
  the theoretical treatment in a certain way.
  The overall density of of the WL electrons is then obviously
  determined by $\overline{N_{\bf k}}$ and
$k_m$.

In the above specified single-electron basis the full Hamiltonian consists
of  free electron Hamiltonian in the bound and WL states, $H_e$,  free phonon
 Hamiltonian,
$H_{ph}$, and the electron-phonon coupling $H_1$ between them:
\bea
H&=&H_e+H_{ph}, \nonumber \\
H_e&=&\sum_{i,n}E_nc^+_{i,n}c_{i,n}+\sum_{\bf k}
E_{{\bf k}}c^+_{{\bf k}}c_{{\bf k}}, \nonumber \\
H_{ph}&=&\sum_{\bf q}E_{LO}b^+_{\bf q}b_{\bf q}, \nonumber  \\
H_1&=&\sum_{\lambda, \mu,{\bf
q}}A_q\Phi(\lambda,\mu,{\bf q}) (b_{\bf q}-b^+_{-{\bf
q}})c^+_{{\lambda}}c_{{\mu}},
\label{Hamiltonian}
\eea
Here  $\lambda$ and $\mu$ are the indexes of
single electron states, either $(i,n)$, or ${\bf k}$, where
$i$ is the quantum dot number and
$n$ is the electron orbital number. $c$ and $b$ are, respectively,
electron and phonon annihilation operators, $E_{LO}$ is phonon energy.
 The coupling constant is
\cite{Callaway}
\be
A_q=-ieq^{-1}[E_{LO}(\kappa^{-1}_{\infty} -\kappa^{-1}_0]^{1/2}
(2\varepsilon_0V)^{-1/2}.
\ee
Here $\kappa_{\infty}$ and $\kappa_0$ are, respectively, the high frequency
and static dielectric constants, $\varepsilon_0$ is the permittivity
of the free space, $V$ is the volume of the sample and $e$ is the
electronic charge. The form-factor in (\ref{Hamiltonian})
\be \Phi(\lambda,\mu,{\bf q})=\int d^3{\bf r}
\psi^*_{\mu}({\bf r})e^{i{\bf q}{\bf r}}\psi_{\lambda}({\bf
r}).
\ee
modifies the Fr\"ohlich's coupling to the case of quantum dot.
 Because the Fr\"ohlich's coupling does not change
the electron spin, we treat the electronic subsystems with the given
spin separately.
The functions $\psi$ are the single electron orbitals
in the SAQD sample.
 The terms of $H_1$
corresponding the Fr\"ohlich's coupling between
two bound states localized in different quantum dots will be
considered as zero.

\section{Spectral density}
We shall calculate the electronic spectral density for the Hamiltonian
specified above. This basic quantity will be used to compare the theoretical
results with some experimental data in photoluminescence,
photoluminescence excitation and optical absorption experiments.
The electronic spectral density $\sigma_{i,n}(E)$,
$E$ being the energy variable, is related to the
retarded  Green's function $G_{i,n}(E)$ with help of the
formula $\rho_{i,n}(E)=-\frac{1}{\pi}ImG_{i,n}(E)$.
Similarly,  spectral densities and the corresponding Green's
functions $G_{\bf k}(E)$ are introduced for the wetting layer electronic states.

The electronic Green's function can be determined by the corresponding
electronic self-energy, $M_{i,n}(E)$.
 Only the diagonal terms, in the electronic orbital
quantum number $n$, of the
 Green's function and
self-energies, will be considered.
      The reader is referred to the earlier works on the
electron energy relaxation in  individual quantum dots
\cite{tsuchiya,1997,se,1998}.
The  bound states electronic self-energy in the self-consistent
Born approximation  reads:
\begin{eqnarray}
\label{discreteM}
M_{i,n}(E)&=&\sum_m \alpha_{n,m}\biggl\{ \frac{1-N_{i,m}+\nu_{LO}}{
E-E_m-E_{LO}-M_{i,m}(E-E_{LO})} \nn  \\
&+&\frac{N_{i,m}+\nu_{LO}}{E-E_m+E_{LO}-M_{i,m}(E+E_{LO})} \biggr\} \nn \\
&+&\sum_{\bf k}\alpha_{n,{\bf k}}\biggl\{ \frac{1-N_{\bf k}+\nu_{LO}}{
E-E_{\bf k}-E_{LO}-M_{\bf k}(E-E_{LO})}  \nn  \\
&+&\frac{N_{\bf k}+\nu_{LO}}{E-E_{\bf k}+E_{LO}-M_{\bf k}(E+E_{LO})}
 \biggr\},   \nn   \\
\end{eqnarray}
while for the wetting layer electrons we have:
\begin{eqnarray}
\label{continuousM}
M_{\bf k}(E) & =&\sum_m \alpha_{{\bf k},{\bf k'}}\biggl\{
\frac{1-N_{\bf k'}+\nu_{LO}}{
E-E_{\bf k'}-E_{LO}-M_{\bf k'}(E-E_{LO})} \nn  \\
& +&\frac{N_{\bf k'}+\nu_{LO}}{E-E_{\bf k'}+E_{LO}-M_{\bf
k'}(E+E_{LO})} \biggr\} \nn \\
& +&\sum_{i,s}\alpha_{{\bf k},{i,s}}\biggl\{ \frac{1-N_{i,s}+\nu_{LO}}{
E-E_{s}-E_{LO}-M_{i,s}(E-E_{LO})}  \nn  \\
& +&\frac{N_{i,s}+\nu_{LO}}{E-E_{s}+E_{LO}-M_{i,s}(E+E_{LO})}
 \biggr\}.
\end{eqnarray}
Here $i$ is the number of quantum dot, while $n$ and $s$ are the
numbers of electron bound state orbitals in the given dot. $N_{i,m}$ is the number
of electrons in the $m$-th state if $i$-th quantum dot, $N_{\bf k}$ is the number
of electrons in the ${\bf k}$-th state of the wetting-layer states and $\nu_{LO}$ is
the population of the optical phonon states of the sample, given here by the
Bose-Einstein distribution function.
 In these
equations for the electronic self-energy, the constants $\alpha$ are
generally given by \be \alpha_{\lambda,\mu}=\sum_{\bf q}\mid A_q \mid^2
\mid \Phi(\mu,\lambda,{\bf q})\mid^2, \label{alpha} \ee where the
summation extends over all optical phonon wave vectors ${\bf q}$. The
above equations for the electronic self-energy are self-consistent.
Generally, they  correspond to including an infinite number of terms of
the self-energy expansion in the powers of $H_1$. The use of the
self-consistency, corresponding to the inclusion of the multiple-phonon
scattering,
 proved important in the case of the electron-energy
relaxation among the discrete energy levels of an electron in quantum dots
\cite{jarelax,tsuchiya}
and would be important when considering multiple bound states
in a single dot, coupled via $H_1$.
Those terms in the equations (\ref{discreteM},\ref{continuousM}),
corresponding to the scattering between one
bound state and one extended state, and those, corresponding to the scattering
of the electron between two extended states, will be  included in the
bare Born approximation only.

Let us  now specify the constants $\alpha$ in the three specific
cases of the choice of the electronic states.
In the case when both $\lambda$ and ${\mu}$ denote the states localized in
the same quantum dot, $\lambda=(i,n)$, $\mu=(i,m)$, we find that
the corresponding constant
$\alpha_{n,m}=\alpha_{m,n}$ in the equation (\ref{alpha})
does not depend on the
quantum dot site index $i$. This is because
 the form-factor  depends on the site index $i$ in the following way:
\be
\Phi((i,m),(i,n),{\bf q})=e^{i{\bf qr}}\Psi^{(irr)}(m,n,{\bf q}),
\ee
where the irreducible part of the form-factor is defined as
\be
\Psi^{(irr)}(m,n,{\bf q})=\int d^3{\bf r}\psi^*_{(0,n)}({\bf r}) e^{i{\bf qr}}
\psi_{(0,m)}({\bf r}).
\ee
Here the site index $0$ means the position of the quantum
dot at the origin of coordinates. We therefore drop out the site index $i$
from the constant $\alpha$ completely
for the coupling between the bound states, and we
have:
\be
\alpha_{nm}=\sum_{\bf q}\mid A_q \mid^2 \mid \Psi^{(irr)}(m,n,{\bf q})
\mid^2.
\ee

The constant $\alpha_{n,{\bf k}}=\alpha_{{\bf k},n}$, which characterizes the
mutual coupling of the bound and extended states, is determined with
help of the form-factor
\be
\Phi((i,n),{\bf k},{\bf q})=V^{-1/2}\int d^3{\bf r} e^{i{\bf kr}}
e^{i{\bf qr}}\psi_m({\bf r}),
\ee
$V$ is the volume of the sample. When the electron wave vector
${\bf k}$ is put zero, the latter integral
is the Fourier  transform of the electronic wave function.
Because this wave function is localized in the cube of the
lateral dimension $d$, the Fourier transform will be a function of ${\bf q}$,
the width of which in the ${\bf q}$-space will be about $\pi/d$. Having confined
 our
space of available extended states to the plane waves the ${\bf k}$-vectors of which
fulfill the condition $\mid {\bf k} \mid \ll \pi/d$, we
can expect that in the limit of  ${\bf k}$-vector going to zero, the magnitude of
${\bf k}$
in the form-factor $ \Phi((i,n),{\bf k},{\bf q})$ will not have any important
influence on the value of the corresponding constant $\alpha$. In practice
we will assume that $\mid {\bf k} \mid \le k_m$, $2\pi/k_m=1.5\times d$.
We get:
\be
\alpha_{n,{\bf k}}\approx \alpha_{n,{\bf k}=0}=\sum
\sum_{\bf q}\mid A_q \mid ^2 \mid \Phi^{(irr)}({\bf k}=0,n,{\bf q})\mid^2,
\ee
where
\be
\Phi^{(irr)}({\bf k}=0,n,{\bf q})=V^{-1/2}\int d^3{\bf r}
e^{i{\bf qr}}\psi_{0,n}({\bf r}).
\ee
It is then straightforward to get
$\alpha_{n,{\bf k}}={B_n}/{V}$,
where the constant $B_n$ is
\be
B_n=\frac{e^2E_{LO}(\kappa^{-1}_{\infty}-\kappa^{-1}_0)}{16\pi^3\varepsilon_0}
\mid \int d^3 {\bf q} e^{i{\bf qr}}\psi_{0n}({\bf r}) \mid^2.
\ee

In the equation (\ref{continuousM}),
the term
which introduces the coupling between the extended states, is (we just rewrite the original
Fr\"ohlich's coupling)
\be
\alpha_{{\bf k},{\bf k}'}\equiv \sum_{\bf q} \mid A_q \mid^2 \mid
\Phi({\bf k},{\bf k}',{\bf q})\mid^2,
\ee
with $\Phi({\bf k},{\bf k}',{\bf q})=\delta_{{\bf k},{\bf k}',+{\bf q}}$.
Then
$\alpha_{{\bf k}',{\bf k}}=C_{{\bf k}'-{\bf k}}/V$,
and
\be
C_{{\bf k}'-{\bf k}}=\frac{e^2E_{LO}(\kappa^{-1}_{\infty}-\kappa^{-1}_0)}{
2\varepsilon_0\mid {\bf k}'-{\bf k}\mid^2}=\frac{p}{\mid {\bf k}'-{\bf k}\mid^2},
\ee
defining the quantity $p$ by this formula.

\section{The influence of the wetting layer states}

We shall now consider  the influence of the wetting layer states on
the spectral density in the sub-wetting layer region of electronic energy.
We have set the bottom of the wetting layer states
 at the energy $E=0$. The electronic states localized in
 quantum dots are then  at $E<0$.

Inspecting the equations (\ref{discreteM}) and
(\ref{continuousM}) we see that  only some of the terms at the right hand side
give
singularities of the self-energy in the sub-wetting layer region,
in the lowest
order of iteration of these equations.
These terms are only the following ones:  the term corresponding to the
last fraction in
equation (\ref{discreteM}) and similarly
the term containing the second fraction in equation
(\ref{continuousM}).
These terms bring
phonon satellites into the electronic spectral density at the sub-WL region.
Also,
these terms give a contribution, which is proportional to the electronic
population of the
wetting-layer states, which property appears to be connected with the
presence of the discrete level
broadening in the sub-WL  region \cite{Nakaema}.

 The experiments,
with which we  wish to compare our results,
 are usually performed at
10\,K.  Therefore, we take $\nu_{LO}=0$ for the population of the phonons.
        So that, unless
the temperature of the lattice optical phonons is increased substantially,
the above self-energy terms are nonzero only
when there is a nonzero population of the wetting layer electron states.
In the present numerical estimates we shall
 assume a rather large population in the wetting-layer states.
 The mechanism of achieving such a WL population is not discussed here
in any detail. The large population of the WL states corresponds
to the numerical estimates in the experimental papers \cite{Nakaema}.
\begin{figure}
\epsfysize=7cm
\begin{center}
\epsffile{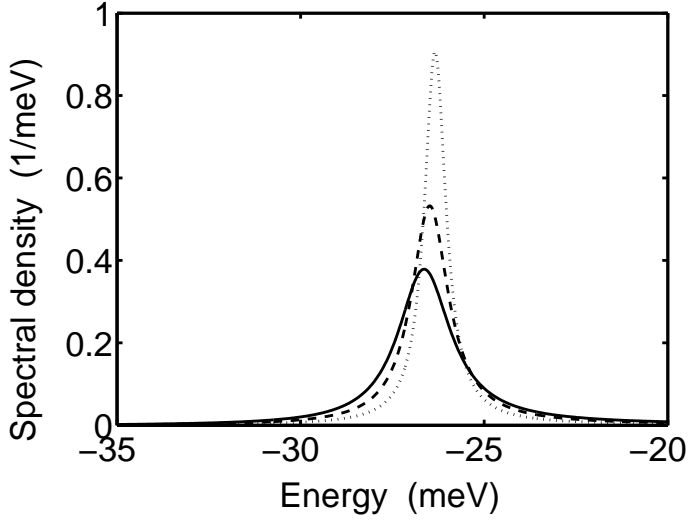}
\end{center}
\caption{Spectral density $\sigma_n(E)$ of the bound state of a single
quantum dot interacting with the wetting-layer states at the
WL polulation: $\overline{N_{\bf k}}= $0.25
(dotted line), 0.5 (dashed) and 0.75 (full).}
\label{bound1}
\end{figure}

\subsection{A SINGLE BOUND STATE}
Let us see the influence of the wetting layer states on the
spectral density of a single bound electronic state.
Let us assume the simple case, when there is only a single
bound state in every quantum dot, which we denote as the state $n=2$,
the energy of which will be denoted as $E_2$ and   $E_2=-E_{LO}+\eta$.
The parameter $\eta$ determining the position of the bound state
energy level is chosen to be 10\,meV. In this work we use
the material parameters of Gallium Arsenide.       Namely, the electron
relative effective mass is   0.067, the static dielectric constant
is $\kappa_0=$12.91 (SI units) and the high frequency dielectric constant is
$\kappa_{\infty}=$10.91. The bulk crystal phonon energy in GaAs is 36.2\,meV.
The wave function of this state
will be assumed to be
that of an electron in one of the states of the triply degenerate first
excited  energy level
of an infinitely deep cubic quantum dot with the lateral size $d$.
Namely, this wave function is
\be
\psi_{02} ({\bf r})=
(2/d)^{3/2} cos(\pi x/d) cos(\pi y/d) sin(2\pi z/d)
\ee
(reminding that the index $0$ means the wave function
placed at the origin of coordinates). The natural line width of the spectral
density of the bound state is taken here to be $\Delta=0.05\,$meV.
The same natural line-width is assumed for the
states in the wetting-layer.
This means that the spectral density
of the unperturbed bound state
would be a Lorenzian, having the halfwidth of 0.05\,meV,  being placed
at the energy $-E_{LO}+10\,$meV.
The electronic self-energy term, expressed in the Born approximation to the coupling
between the bound state $\psi_{02}$
and the WL states, and being proportional to the WL population at low
temperatures, is simply
\be
M_n(E)= \sum_{\bf k}\alpha_{n,{\bf k}}\frac{N_{\bf k}}{E-E_{\bf
k}+E_{LO}+i\Delta}.
\label{diM}
\ee
We omit the index of the quantum dot here. As said above,
we assume here that the population of the WL states $N_{\bf k}$
is constant for
all $\overline{N_{\bf k}}$. Under the above made assumption,
according to which the wetting-layer states ${\bf k}$-vector magnitude
is restricted to be not larger than
  $k_m$, this means that the average density of the electrons (of both spins)
  is about $1.5\times$10$^{23}$\,m$^{-3}$ for $\overline{N_{\bf k}}=0.5$.
  We assume that the electrons occupy only  the WL states.

Reminding the above introduced approximation, according to which the
wavevector dependence of the constant
$\alpha_{n,{\bf k}}$ is neglected, the integral in the equation (\ref{diM})
can be integrated analytically
\cite{Dwight}.
The result of the numerical evaluation  of the electronic
spectral density $\sigma_n(E)$ of the electronic state $n=2$,
bound in a quantum dot,  is presented in Fig.\ \ref{bound1}
for three values of the population of
the wetting layer states. The spectral peak
of the bound state is broadened, from the unperturbed value of the full width
at half maximum being 0.1\,meV, to the value of about 2\,meV. This spectral
width is of the same order of magnitude as it may be found in the experimental
papers (see e. g. \cite{Nakaema}. The Figure\ \ref{bound1}
shows the trend of the discrete state spectral density to broaden with the
increase of the wetting layer states population. This behaviour of the
bound state  line-width
appears to be in accord
with the experiment \cite{Nakaema}.
The purpose of the present study is to pay attention
to the main features of the effects under consideration. From this reason
we use the material parameters of GaAs, as a
typical material among the polar semiconductors.
The quantum dot samples studied in experiment are often the InGaAs quantum dots,
 and the corresponding
 weting layer, both
 grown on the substrate of GaAs. Although the parameters characterizing
 InGaAs are rather similar to those of GaAs, we nevertheless do not compare our results
 quantitatively with the experiment. In the case of the
 bound state energy level broadenig,  the reader is referred to the paper
 \cite{Nakaema}, namely to the Figure 3 in that paper, for the experimental data on the
 level broadening as it depends indirectly on the WL electron density.  Let us emphasize that
 in the experimental works, with which we compare our results, the optical spectra
 are obtained in such a way, that the they are measured in the limit
 of a single dot measurement and therefore the inhomogeneous broadening of the optical spectra
 is avoided. In this sense, we
 can therefore compare the experimental data with our "homogeneously" broadened
 theoretical shape.

Besides the broadening of the discrete electronic level, the coupling
of the bound state to the WL states tends to give a weak
continuous background. Quantitatively, this background appears
rather weak at the presently
used values of the input parameters of the material, quantum dot size and shape,
and at the present assumption about the populated states in the
wetting-layer continuum.
We may conclude that, the interaction of the bound state in the dot with
the wetting layer states, under rather high population per quantum mechanical
state, leads mainly to the bound state level broadening
in the sub-wetting layer region.

\subsection{PHONON SATELLITE OF THE WETTING LAYER STATES}

We do not pay attention
to the changes of the spectral density in the WL region, which  come from the
coupling of the WL states to the bound states.
In the absence of the quantum dots in the sample, the electron-phonon
interaction between the wetting-layer states  leads, as it is well known
\cite{Mahan},
to the formation of the
polaron state of an electron, in which the
 electron is wrapped by the cloud of the optical phonons. The depth
of the effective electronic polaron potential
hole in the crystal of GaAs may be about 2.5\,meV \cite{Callaway}.
In this case the spread of the polaron cloud may be estimated as
several tens of nanometers.
When the density of the quantum dots in the sample is large enough, so that the
inter-dot separation is comparable to the size of the polaron,
the existence of the polaron states may be not well established. In an overall
agreement with the current
experiments \cite{Toda,Nakaema} we shall assume the density of the quantum dots in the sample
to be 8$\times 10^{21}$\,m$^{-3}$.
Realizing that the polaronic spectral density peak would be only  about
several meV below the low-energy edge of the wetting-layer states in GaAs,
and having in mind
the intention to study a mechanism leading to the appearance of a broad
continuous spectral density in the broad  range of energy
from about $E=-E_{LO}$ until the wetting-layer edge $E=0$,
we shall leave the question of   the polaron effect of the carriers in the WL states
open and give our attention
to that term in the equation (\ref{continuousM}), which is at $T=0$ proportional
to the electronic population of the wetting layer $N_{\bf k}$.

\begin{figure}
\epsfysize=7cm
\begin{center}
\epsffile{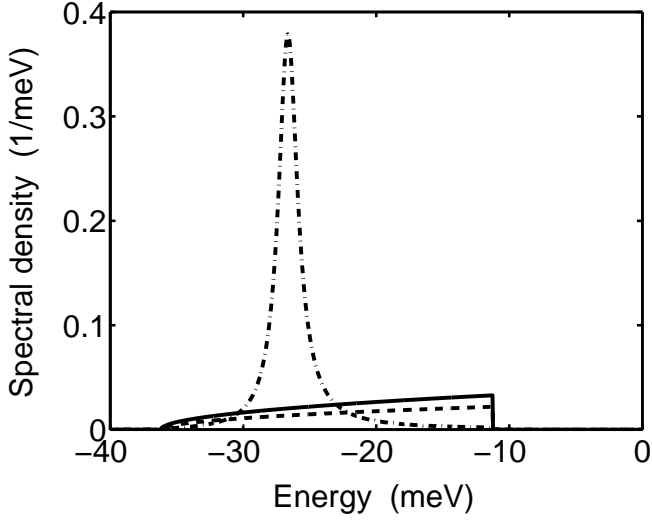}
\end{center}
\caption{The full and the dash-dot lines  give, respectively,
 the spectral densities of the  phonon satellite
of the wetting-layer states $\sigma_{tot}(E)$, related to
a single quantum dot, and the spectral density $\sigma_n(E)$
of the discrete state of a single dot, both computed for the wetting-layer
population $\overline{N_{\bf k}}=0.75$.
The dashed line is  the spectral density
of the phonon sattelite of the WL states   $\sigma_{tot}(E)$
computed for the  WL states
population $\overline{N_{\bf k}}=0.5$. }
\label{boundacont}
\end{figure}

That term of the self-energy of the wetting-layer state ${\bf k}$, which is
proportional to the wetting-layer population, will be
 in the lowest order approximation:
\be
M_{\bf k}(E)=\sum_{{\bf k}'}
\alpha_{{\bf k},{{\bf k}'}}
\frac{N_{{\bf k}'}}
{
E-E_{{\bf k}'}  +  E_{LO}+i\Delta
}.
\ee
Substituting for $\alpha_{ {\bf k} , {{\bf k}'} }$ we get
\be
M_{\bf k}(E)=\frac{1}{(2\pi)^3}\int\!\!\int\!\!\int
d^3{\bf k}'\frac{p}{\mid {\bf k}'-{\bf k}\mid^2}
\frac{N_{{\bf k}'}}{E+E_{LO}-E_{{\bf k}'}+i\Delta}.
\ee
The integration is extended over the range of ${\bf k}<k_m$.
Changing the integration variable from
${\bf k'}$ to ${\bf w}={\bf k'}-{\bf k}$, we get
\be
M_{\bf k}(E)=\frac{1}{(2\pi)^3}\int\!\!\int\!\!\int d^3{\bf w}
\frac{p}{\mid {\bf w}\mid ^2}
\frac{N_{{\bf w}+{\bf k}}}
{E+E_{LO}-E_{{\bf w}+{\bf k}}+i\Delta}.
\label{simple}
\ee
When going to the spherical coordinates, the factor
$\mid {\bf w}\mid^2$ in the denominator cancels with the factor
coming from the Jacobian of the transformation.
In the rest of the integrand we shall make
the approximation of "forward scattering", which will consist in
neglecting the scattering vector ${\bf w}$ in $N_{{\bf w}+{\bf k}}$ and in
$E_{{\bf w}+{\bf k}}$. The integration limits obviously must obey simultaneously
the complicated
condition of
$\mid {\bf k} \mid<k_m$ and  $\mid {{\bf w}+{\bf k}}\mid <k_m$.
We shall substitute this condition by the condition
$\mid {\bf w}\mid <k_m$ which is obviously exact in the limit of ${\bf k}=0$.
The preliminary detailed numerical evaluation of the integral shows, that the simplified
evaluation of the integral in (\ref{simple}) is plausible.

The above introduced approximations allow for obtaining
the following simple approximate analytical form of the self-energy, namely,
\be
M_{\bf k}(E)=\frac{\gamma}{E+E_{LO}-E_{\bf k}+i\Delta},
\ee
in which $\gamma=pk_m\overline{N_{\bf k}}/(2\pi^2)$. The distribution
of the electrons in all the wetting-layer states ${\bf k}$ is constant and equal to
$\overline{N_{\bf k}}$.

The Green's function of the state ${\bf k}$ in the wetting layer then is:
\be
G_{\bf k}(E)=\frac{1}{E-E_{\bf k}-
\frac{\gamma}{E+E_{LO}-E_{\bf k}+i\Delta}+i\Delta}.
\ee
Looking for the approximate single pole behaviour of this Green's function in the
region of energy $E\approx E_{LO}$, we find the following approximate
expression of the wetting-layer electron Green's function:
\be
G_{\bf k}(E)=\frac{\gamma/E_{LO}^2}{E+E_{LO}-E_{\bf k}+i\Delta}.
\ee
Summing up all the satellite spectral densities of the individual WL
states ${\bf k}$ within the
range of ${\bf k}<k_m$ gives the total spectral density $\sigma_{tot}(E)$
of all the extended states.
The total spectral density of the phonon satellite of the WL states,
related to a single quantum dot, is displayed in the
Fig.\ \ref{boundacont} for the values 0.5 and 0.75 of the population
$\overline{N_{\bf k}}$. In the same graph the single dot bound state spectral
density for $\overline{N_{\bf k}}=0.75$ is displayed for the purpose
of comparing the spectral density of the single dot bound state spectral line
with the spectral density of the phonon satellite per one quantum dot.

The total spectral density of the WL satellite, $\sigma_{tot}(E)$, is proportional
to the first power of the WL population $\overline{N_{\bf k}}$, while the integral of the
discrete state $\sigma_n(E)$ is constant and equal one.
Therefore, increasing the WL population leads
to a relative weakening of the discrete state spectral density with respect to the
$\sigma_{tot}(E)$. A similar trend is observed in the experimental paper \cite{Nakaema},
in which the discrete peaks  in the sub-wetting layer region of the spectra
appear to weaken with respect to the continuous background signal,
when the WL population increases.

In the present model of the SAQD sample, the set of the WL states included
in the theory is identical with
the set of states which are
populated with nonzero $N_{\bf k}$. Only these states contribute then
to the formation of the phonon satellite of the WL states. This may lead us to expect,
that the experimental spectra of the sub-WL optical response
 might be influenced also by
the distribution function of the population of the WL states.

In the present approach the WL satellite continuum is found in the interval of energy
$(-E_{LO},0)$, while in the experiment the continuous background may be
observed even at the lower energy side of this  energy interval.
We expect that our model may give the extended
range of the continuum  upon going to higher orders in the iterative solution of the
equations
(\ref{discreteM},\ref{continuousM}).

The numerical estimate of the dimensionless
quantity $\gamma/E_{LO}^2$ gives the value of about
2.8$\times 10^{-2}$. The spectral density of the wetting layer satellite
should thus be
about 36 times smaller than that of the WL spectral density itself.
 This number is not easy to compare
with the available
experimental data, see e. g. \cite{Heitzmulti},  because, e. g., we have to be aware
of the fact that the optical spectra
may be not simply proportional to the spectral densities and  also, because our
model, assuming the three dimensional wetting layer, may not allow
for such a direct comparison of these spectral densities.

\section{Conclusions}
The present numerical estimate shows that the Fr\"ohlich's mechanism of the
electron-phonon coupling in the self-assembled quantum dot samples can contribute
considerably to
 the spectral features observed in the PLE experiments, providing that the
 population
of the electrons in the wetting-layer states is sufficiently large.
The optical signal in the sub-wetting layer region of the
excitation energy can be expected
to have a twofold origin. First, the energy levels of the bound states are broadened
by the coupling to the populated WL states. Second, there is a continuous
background in the sub-WL
region, which is ascribed to the low-energy phonon sattelite of the wetting-layer
states and which intensity increases linearly with the WL population.

The present conclusions, saying that the electron-LO-phonon coupling can provide
the explanation of the continuous features in the optical spectra, do not mean an
 exclusion of
other contributions, like  the direct
carrier-carrier coupling. These effects remain
to be considered in detail, together with the influence of the polaron states and
 the dimensionality of the WL states.

\acknowledgements{The work was supported by the grants IAA1010113,
OCP5.20, RN19982003014  and by the project AVOZ1-010-914.}

\end{document}